\documentclass[11pt]{article}
\usepackage{epsfig}
\usepackage{graphicx}
\usepackage{amsmath}
\usepackage{amssymb}
\usepackage{amsxtra}
\usepackage{amsfonts,amsthm}
\usepackage{subfigure}
\usepackage[rflt]{floatflt}
\usepackage{color}
\usepackage[T1]{fontenc}
\usepackage[latin9]{inputenc}
\def\Journal#1#2#3#4{{#1} {\bf #2} (#4) #3 }

\def\PRL{ Phys. Rev. Lett.}

\def\IJMPA{{ Int. J. Mod. Phys.}  A}

\def\BWP{ Bled Workshops in Physics}
\def\s{{\,\rm s}}
\def\g{{\,\rm g}}
\def\eV{\,{\rm eV}}

\def\({\left(}
\def\){\right)}
\def\cm{{\,\rm cm}}
\def\km{{\,\rm km}}

\def\beq{\begin{equation}}
\def\eeq{\end{equation}}
\def\lsim{\mathrel{\rlap{\lower4pt\hbox{\hskip1pt$\sim$}}
    \raise1pt\hbox{$<$}}}         

\def\gsim{\mathrel{\rlap{\lower4pt\hbox{\hskip1pt$\sim$}}
    \raise1pt\hbox{$>$}}}         
\title{Some potential problems \\ of OHe composite dark matter}
\author{J.R. Cudell$^{a,}$\footnote{JR.Cudell@ulg.ac.be}\\
M.~Khlopov $^{b,}$\footnote{Khlopov@apc.univ-paris7.fr}\\
Q. Wallemacq
$^{a,}$\footnote{Quentin.Wallemacq@ulg.ac.be}}
\begin{document}
\maketitle
\begin{flushleft}{{\it $^{a}$ \small IFPA, D\'ep. AGO, Universit\'e de Li\`ege, Sart Tilman, 4000 Li\`ege, Belgium\\\
      $^{b}$ \small National Research Nuclear University "MEPHI" (Moscow Engineering Physics
Institute), 115409 Moscow, Russia \\
Centre for Cosmoparticle Physics ``Cosmion'' 115409 Moscow, Russia
\\ APC laboratory 10, rue Alice Domon et Lonie Duquet
75205 Paris Cedex 13, France}}
\end{flushleft}
\begin{abstract}
Among composite-dark-matter scenarios, one of the simplest and most predictive is 
that  of O-helium (OHe) dark atoms, in which a lepton-like doubly charged particle O$^{--}$ is bound 
with a primordial helium nucleus, and is the main constituent of dark matter. This model liberates
the physics of dark matter from many unknown features of new physics, and it demands
a deep understanding of the details of known nuclear and atomic physics, which are still somewhat
unclear in the case of nuclear interacting ``atomic'' shells. So far the model has relied on the
 dominance of elastic scattering of OHe with the matter. In view of the uncertainty in our understanding 
 of OHe interaction with nuclei we  study the opposite scenario, in which inelastic nuclear reactions dominate
the OHe interactions with nuclei. We show that in this case all the OHe atoms bind with extra He nuclei, 
forming doubly charged O-beryllium ions, which behave like anomalous helium, causing potential problems 
with overabundance of anomalous isotopes in terrestrial matter.
\end{abstract}

\section{Introduction}
Direct searches for dark matter have produced surprising results. Since the DAMA collaboration observed a
signal, several other collaborations seem to confirm an observation, while many others
clearly rule out any detection. The current
experimental situation is reviewed in \cite{DAMAtalk}. This apparent
contradiction comes from the analysis of the data under the assumption that nuclear
recoil is the source of the signal.

Starting from 2006 it was proposed \cite{KK1,invention,spectro,DMRev,DDMRev} that the
signal may be due to a different source: if dark matter can bind to
normal matter, the observations could come from radiative capture of thermalized dark
matter, and could depend on the detector composition and temperature. This scenario naturally comes
 from the consideration of composite dark matter. Indeed, one can imagine that
dark matter is the result of the existence
of heavy negatively charged particles that bind to primordial nuclei.

Cosmological considerations imply that such candidates for dark matter should consist of
negatively doubly-charged heavy ($\sim 1$ TeV) particles, which we call O$^{--}$, coupled
to primordial helium. Lepton-like technibaryons, technileptons, AC-leptons or clusters of
three heavy anti-U-quarks of 4th or 5th generation with strongly suppressed hadronic
interactions are examples of such O$^{--}$ particles (see \cite{invention,spectro,DMRev,DDMRev} for
a review and for references).

 It was first assumed that the effective
potential between OHe and a normal nucleus would have a barrier, preventing He and/or
O$^{--}$ from falling
into the nucleus, allowing only one bound state, and diminishing considerably the
interactions of OHe. Under these conditions elastic collisions dominate in OHe
interactions with matter, and lead to a successful OHe scenario.
The cosmological and astrophysical effects of such composite dark matter (dark atoms of
OHe) are dominantly related to the helium shell of OHe and involve only one parameter
of new physics $-$ the mass of O$^{--}$. The positive results of the DAMA/NaI and
DAMA/LIBRA
experiments are explained by the annual modulations of the rate of radiative capture of OHe
by sodium nuclei. Such radiative capture is possible only for intermediate-mass nuclei:
this explains the negative results of the XENON100 experiment. The rate of this capture is
proportional to the temperature: this leads to a suppression of this effect in cryogenic
detectors, such as CDMS. OHe collisions in the central part of the Galaxy lead to OHe
excitations, and de-excitations with pair production in E0 transitions can explain the
excess of the positron-annihilation line, observed by INTEGRAL in the galactic bulge \cite{DMRev,DDMRev,CDM,KK2,KMS,CKW}.
In a two-component dark atom model, based on Walking Technicolor, a sparse WIMP-like component of atom-like 
state, made of positive and negative doubly charged techniparticles, is present together with the dominant OHe dark atom and the decays of 
doubly positive charged techniparticles to pairs of same-sign leptons can explain the excess of 
high-energy cosmic-ray positrons, found in PAMELA and AMS02 experiments \cite{laletin}.

These astroparticle data can be fitted, avoiding many astrophysical uncertainties of WIMP
models, for a mass of O$^{--}$ $\sim 1$ TeV, which stimulates searches for stable doubly
charged lepton-like particles at the LHC as a test of the composite-dark-matter scenario.

In this paper, we want to explore the opposite scenarion, in which OHe dark matter interacts strongly
with normal matter: OHe is neutral, but a priori it has an unshielded nuclear attraction
to matter nuclei.
We first study some effects of inelastic collisions of OHe in the early Universe and in the terrestrial matter and find that such collisions strongly 
increase the formation of charged nuclear species with O$^{--}$ bound in them. Recombination of such charged species with electrons 
(even if it is partial) leads to the formation of atoms (or ions) of anomalous isotopes of helium and heavier elements. The atomic size of such
 atoms (or ions) of anomalous isotopes strongly suppresses their mobility in the terrestrial matter, making them stop near the surface, 
 where anomalous superheavy nuclei are strongly constrained by the experimental searches.
In Section \ref{BBN} we study effect of inelastic processes during the period of Big Bang Nucleosynthesis and show that if these processes are 
not suppressed all the OHe atoms capture additional He nuclei, forming a doubly charged ion of O-beryllium (OBe). In Section \ref{mobility} 
we briefly examine the problems of an OBe-dominated universe and show that, because the mobility of the anomalous 
isotopes is greatly suppressed even if they recombine with only one electron, their drift to the center of the Earth is strongly slowed down, and  
their abundance increases near the terrestrial surface and in the World Ocean with the danger of their overabundance. We stress the importance 
of solving the open questions of OHe nuclear physics in the Conclusion.

\section{Inelastic processes with OHe in the early Universe}
\label{BBN}
As soon as all the OHe is formed in the early Universe, inelastic processes between OHe and OHe itself and between OHe and the primordial He 
take place and start consuming the available OHe.  The two relevant reactions are:
\begin{eqnarray}
\mathrm{OHe}+\mathrm{OHe}&\rightarrow &\mathrm{O_{2}Be}\label{eq1} \\
\mathrm{OHe}+\mathrm{He}&\rightarrow &\mathrm{OBe}\label{eq2}
\end{eqnarray}
Note that in these reactions  the addition of a He nucleus to the bound OHe system will result in merging the two He nuclei into $^8$Be, since in 
the presence of O$^{--}$, $^8$Be becomes stable:  we calculated, as in Ref. \cite{CKW3}, that the energy of OBe is 2.9 MeV smaller than 
that of OHe+He.
The temperature $T_0$ at which OHe forms depends on its binding energy, which has been accurately evaluated as 1.175 MeV in Ref. 
\cite{CKW3}, and corresponds approximately to $T_0=50$ keV. As the cosmological time $t$ is related to the temperature through 
$t(\mathrm{s})\simeq\frac{1}{T^2(\mathrm{MeV})}$, processes \eqref{eq1} and \eqref{eq2} start at a time $t_{0}\simeq\frac{1}{0.05^2}
=400$ s after the Big Bang and continue until helium freezes out at $t_{*}\simeq 10~\mathrm{min} = 600$ s.

During these 200 s, the OHe atoms are consumed at a rate:
\begin{equation}
\frac{\mathrm{d}n_{\mathrm{OHe}}}{\mathrm{d}t}=-3Hn_{\mathrm{OHe}} - n_{\mathrm{OHe}}^2\sigma_1 v_1 - n_{\mathrm{OHe}}
n_{\mathrm{He}}\sigma_2 v_2,
\label{eq3}
\end{equation}
where $n_{\mathrm{OHe}}$ and $n_{\mathrm{He}}$ are the number densities of OHe and He, $H=\frac{1}{2t}$ is the expansion rate of the 
Universe during the radiation-dominated era, $\sigma_1$ and $\sigma_2$ are the cross sections of processes \eqref{eq1} and \eqref{eq2} 
respectively and $v_1$ and $v_2$ are the OHe-OHe and OHe-He mean relative velocities. The first term in the right-hand side of equation 
\eqref{eq3} corresponds to the dilution in an expanding universe. The number of helium nuclei per comoving volume is assumed to be unaffected 
by reaction \eqref{eq2} since the abundance of helium is more than an order of magnitude higher than that of OHe, so that the only effect on 
$n_{\mathrm{He}}$ is due to the expansion:
\begin{equation}
\frac{\mathrm{d}n_{\mathrm{He}}}{\mathrm{d}t}=-3Hn_{\mathrm{He}},
\label{eq4}
\end{equation}
from which it follows that:
\begin{equation}
n_{\mathrm{He}}(t)=n_{\mathrm{He}}^0\left(\frac{t_0}{t}\right)^{3/2},
\label{eq5}
\end{equation}
where $n_{\mathrm{He}}^0$ is the number density of He at $t=t_0$ (In the following, we shall use a superscript $0$
to denote quantities taken at the time of the decoupling of OHe, $t=t_0$).

To take into account the effect of the expansion and calculate the decrease of the fraction of free OHe atoms due to their inelastic reactions, we 
study the ratio $f$ of the number density of OHe atoms to the number desity of He nuclei, $f=\frac{n_{\mathrm{OHe}}}{n_{\mathrm{He}}}$. 
From \eqref{eq3} and \eqref{eq4}, its evolution is given by:

\begin{equation}
\frac{\mathrm{d}f}{\mathrm{d}t}=-n_{\mathrm{He}}f\left(\sigma_1 v_1 f+\sigma_2 v_2\right)
\label{6}
\end{equation}
The capture cross sections $\sigma_1$ and $\sigma_2$ are of the order of the geometrical ones:
\begin{eqnarray}
\sigma_1&\approx&4\pi \left(2r_{\mathrm{OHe}}\right)^2,\label{eq7} \\
\sigma_2&\approx&4\pi \left(r_{\mathrm{OHe}}+r_{\mathrm{He}}\right)^2,\label{eq8}
\end{eqnarray}
where $r_{\mathrm{OHe}}$  is the Bohr radius of an OHe atom and $r_{\mathrm{He}}$ is the radius of a He nucleus. As both of them are 
approximately equal to 2 fm, $\sigma_1 \approx \sigma_2 \approx 64\pi ~10^{-26}$ cm$^2$. As the OHe and He species are in thermal 
equilibrium with the plasma at temperature $T$, the mean relative velocities $v_1$ and $v_2$ are obtained from the Maxwell-Boltzmann 
velocity distributions of OHe and He and are given by:

\begin{eqnarray}
v_1&=&\sqrt{\frac{8T}{\pi \mu_1}},\label{eq9} \\
v_2&=&\sqrt{\frac{8T}{\pi \mu_2}},\label{eq10}
\end{eqnarray}
where $\mu_1=m_{\mathrm{OHe}}/2$ and $\mu_2 \simeq m_{\mathrm{He}}$ are the reduced masses of the OHe-OHe and OHe-He systems. 
$m_{\mathrm{OHe}}=1000$ GeV is the mass of an OHe atom, and $m_{\mathrm{He}}=3.73$ GeV that of a He nucleus. Given the time 
dependence of the temperature during the radiation-dominated era, $Tt^{1/2}=T_0t_0^{1/2}$, one can use it to express the velocities 
\eqref{eq9} and \eqref{eq10} as functions of time and insert the resulting expressions together with \eqref{eq5} in equation \eqref{6} and get:
\begin{equation}
\frac{\mathrm{d}f}{\mathrm{d}t}=-\gamma\frac{1}{t^{7/4}}f\left(\alpha f+\beta \right),
\label{eq11}
\end{equation}
with
\begin{eqnarray}
\alpha &=&\frac{\sigma_1}{\sqrt{\mu_1}},\label{eq12} \\
\beta &=&\frac{\sigma_2}{\sqrt{\mu_2}},\label{eq13} \\
\gamma &=&n_{\mathrm{He}}^0t_0^{7/4}\sqrt{\frac{8T_0}{\pi}}.\label{eq14}
\end{eqnarray}

The solution of \eqref{eq11} corresponding to the initial condition $f(t_0)=f_0$ is given by:
\begin{equation}
f(t)=\frac{\beta f_0}{\exp\left({\frac{4}{3}\beta \gamma \left({t_0^{-3/4}}-{t^{-3/4}}\right)}\right)\left(\alpha f_0+\beta\right)-\alpha}.
\label{eq15}
\end{equation}
The number density of He nuclei at the time of OHe formation, $n_{\mathrm{He}}^0$, can be found from its value $n_{\mathrm{He}}^1$ 
today (In the following, the superscript 1 will denote quantities at the present time). Helium nuclei represent nowadays approximately 10$\%$ of all baryons, which have an energy density $\rho_B^1$ of about 5$\%$ of 
the critical density $\rho_c^1$:  $n_{\mathrm{He}}^1\simeq 0.1n_{\mathrm{B}}^1=0.1\frac{\rho_B^1}{m_p}\simeq 0.1\times 0.05 
\frac{\rho_c^1}{m_p}$, where $m_p$  is the mass of the proton. The present critical density is measured to be $\rho_c^1=5.67\times 
10^{-6} m_p$/cm$^3$, so that $n_{\mathrm{He}}^1\simeq 2.8\times 10^{-8}$ cm$^{-3}$. As it was assumed that the He number density 
was not affected by reaction \eqref{eq2}, the only effect between $t_0$ and now has been a dilution due to the expansion, and hence 
$n_{\mathrm{He}}\propto \frac{1}{a^3}\propto T^3$, where $a$ is the scale factor. Knowing that the temperature of the CMB today is 
$T_1=2.7$ K$=2.33\times 10^{-7}$ keV, this gives $n_{\mathrm{He}}^0=n_{\mathrm{He}}^1\left(\frac{T_0}{T_1}\right)^3\simeq 
2.8\times 10^{-8}\left(\frac{50}{2.33\times 10^{-7}}\right)^3\simeq2.8\times 10^{17}$ cm$^{-3}$.

At the time of OHe formation, all the O$^{--}$ particles were in the form of OHe, i.e. the number density of O$^{--}$ at $t=t_0$, 
$n_{\mathrm{O}}^0$,
was equal to that of OHe,  $n_{\mathrm{OHe}}^0$. Between $t_0$ and today, O$^{--}$ particles may have been bound in different structures, 
but they have not been created or destroyed, so that their number density has only been diluted by the expansion in the same way as that of He 
nuclei, so that the ratio of the number density of O$^{--}$ particles to the number density of He nuclei remains unchanged: 
$\frac{n_{\mathrm{O}}^0}{n_{\mathrm{He}}^0}=\frac{n_{\mathrm{O}}^1}{n_{\mathrm{He}}^1}$. Therefore, the initial fraction $f_0$ of 
OHe atoms can be calculated from present quantities: $f_0=\frac{n_{\mathrm{OHe}}^0}{n_{\mathrm{He}}^0}=\frac{n_{\mathrm{O}}^0}
{n_{\mathrm{He}}^0}=\frac{n_{\mathrm{O}}^1}{n_{\mathrm{He}}^1}$. $n_{\mathrm{O}}^1$ is obtained from the fact that O$^{--}$ 
saturates the dark matter energy density, which represents about 25$\%$ of the critical density: $n_{\mathrm{O}}^1\simeq 
0.25\frac{\rho_c^1}{m_{\mathrm{O}}}\simeq 1.3\times 10^{-9}$ cm$^{-3}$, where $m_{\mathrm{O}}=1$ TeV is the mass of O$^{--}$. 
With the previously calculated value of $n_{\mathrm{He}}^1$, this gives $f_0\simeq 0.05$.

We can now insert the numerical values into Eq.~\ref{eq15} and get the fraction of OHe atoms at the time of helium freeze-out 
 $t=t_{*}=600$ s:
\begin{equation}
f(t_{*})\simeq 5\times 10^{-6133}\ll f_0,
\end{equation}
meaning that no OHe survives reactions \eqref{eq1} and \eqref{eq2}. More precisely, most of the OHe atoms have captured He nuclei via 
process \eqref{eq2} and are now in the form of OBe. Indeed, the majority of the suppression of $f$ comes from the exponential term present in 
\eqref{eq15}, evaluated to be $e^{14127}$. The large argument of the exponential represents the number $N_2$ of reactions 
\eqref{eq2} that happened between $t_0$ and $t_{*}$, per OHe atom:

\begin{align*}
N_2 & = \int_{t_{0}}^{t_{*}} n_{\mathrm{He}}(t)\sigma_2 v_2(t)\mathrm{d}t \\
    & = n_{\mathrm{He}}^0t_0^{3/2}\sigma_2 \sqrt{\frac{8T_0t_0^{1/2}}{\pi \mu_2}}\int_{t_{0}}^{t_{*}}\frac{1}{t^{7/4}}\mathrm{d}t \\
    & =  n_{\mathrm{He}}^0 t_0^{7/4} \sqrt{\frac{8T_0}{\pi}} \frac{\sigma_2}{\sqrt{\mu_2}}\left(-\frac{4}{3}\right)\left(\frac{1}{t_{*}^{3/4}}-\frac{1}{t_{0}^{3/4}}\right) \\
    & = \frac{4}{3} \beta \gamma \left(\frac{1}{t_{0}^{3/4}}-\frac{1}{t_{*}^{3/4}}\right),
\end{align*}
where we have used \eqref{eq5}, \eqref{eq10} and $Tt^{1/2}=T_0t_0^{1/2}$ to pass from the first to the second line and the definitions \eqref{eq13} and \eqref{eq14} for the last line.

Therefore, the realization of the scenario of an OHe Universe implies a very strong suppression of reaction \eqref{eq2}, 
corresponding to $N_2 \ll 1$. Such a suppression needs the development of a strong dipole Coulomb barrier in OHe-He interaction. 
The existence of this barrier and its effect is one of the most important open problems of the OHe model.

\section{Problems of OBe "dark" matter}
\label{mobility} 
Due to Coulomb repulsion further helium capture by OBe is suppressed and one should expect that dark matter is mostly made of doubly charged 
OBe, which recombines with electrons in the period of recombination of helium at the temperature $T_{od}=2 \eV$, before the beginning of 
matter dominance at $T_{RM}=1\eV$. It makes anomalous helium the dominant form of dark matter in this scenario. After recombination 
the OBe gas decouples from the plasma and from radiation and can play the role of a specific Warmer than Cold dark matter, since the adiabatic damping slightly suppresses density fluctuations at scales smaller than the scale of the horizon in the period of He recombination. The total mass
of the OBe gas within the horizon in that period is given by analogy with the case of OHe \cite{KK1,DMRev} by \beq M_{od} =
\frac{T_{RM}}{T_{od}} m_{Pl} (\frac{m_{Pl}}{T_{od}})^2 \approx 2
\cdot 10^{50} \g = 10^{17} M_{\odot}, \label{MEPm}
\eeq where $M_{\odot}$ is the solar mass. 

At momentum values of interest, one finds that elastic cross sections are significantly
enhanced from their geometrical estimate. In the following, we shall use the estimate of 
Ref. \cite{Kaplan}, based on a compilation
of results from general quantum mechanical scattering and from
detailed quantum computations of hydrogen scattering \cite{database}:
\begin{equation}
\sigma=4\pi(\kappa r_0)^2,~\kappa=3-10,\label{sigma}
\end{equation}
with larger values of $\kappa$ at low momentum.

For a size of OBe atoms equal to that of helium $r_0= 3 \cdot 10^{-9} \cm$ and one obtains an elastic scattering cross section
on light elements of the order of $\sigma  \approx 10^{-15} -10^{-14}\cm^2$. It makes this "dark 
matter" follow the ordinary baryonic matter in the process of galaxy formation, and makes it collisional on the scale of galaxies. This causes 
problems with the explanation of the observations of halo shapes \cite{HaloShape}. Presence of OBe in stars can also influence nuclear 
processes, in particular helium burning in the red giants. The processes in stars can lead to the capture by OBe of additional nuclei, thus creating 
anomalous isotopes of elements with higher Z. OBe atoms can also be ionized in the Galaxy, but in the following we shall assume that neutral OBe 
atoms are the dominant part of this "dark matter" on Earth, considering also that slowing down anomalous nuclei in the atmosphere leads to 
ionization and their neutralisation through electron capture.

Falling down on Earth OBe atoms are slowed down and due to the atomic cross section of their collisions have a very low mobility.
After they fall down to the terrestrial surface, the OBe
atoms are further slowed down by their elastic collisions with
matter. They drift, sinking down towards the center of the
Earth with velocity \beq V = \frac{g}{n \sigma v} \leq  2.7 10^{-11} ~\mathrm{cm}/\s\approx 270~ \mathrm{fm}/\s. \label{dif}\eeq 
Here $n$ is the number density of terrestrial atoms, $\sigma v$ is the rate
of atomic collisions, taken at room temperature, and $g=980~ \cm/\s^2$.
We assimilated the crust of the Earth as made of SiO$_2$, and got the number density to be
$n=0.27~ 10^{23}$ molecules/cm$^3$. Using \eqref{sigma}, and taking the geometrical radius to be that of SiO$_2$, 
{\it i.e.} $r_0\approx 2$ \AA,
we obtained $\sigma\geq 4.5~ 10^{-14}$ cm$^2$, and for collisions on SiO$_2$ $v\approx 3~ 10^4 $cm/s.

The OBe abundance in the Earth is determined by the equilibrium between the in-falling and down-drifting fluxes.
The in-falling O-helium flux from dark matter halo is given by \cite{spectro} 
$$
  F=\frac{n_{0}}{8\pi}\cdot |\overline{V_{h}}+\overline{V_{E}}|,
$$
where $V_{h}$ is the speed of the Solar System (220 km/s), $V_{E}$ the speed of the
Earth (29.5 km/s) and $n_{0}=3 \cdot 10^{-4} \cm^{-3}$ is the assumed
local density of OBe dark matter (for an OBe of mass 1 TeV). Furthermore, for simplicity, we didn't take into account the annual modulation of the 
incoming flux and take $|\overline{V_{h}}+\overline{V_{E}}| = u \approx 300 \km/\s$.

The equilibrium concentration of OBe, which is established in the matter consisting of atoms with number density $n$, is given by \cite{spectro}
\begin{equation}
    n_{oE}=\frac{2\pi \cdot F}{V} ,
    \label{noE}
\end{equation}
and the ratio of anomalous helium isotopes to the total amount of SiO$_2$ is given by
\begin{equation}
    r_{oE}=\frac{n_{oE}}{n} = \frac{2\pi \cdot F \sigma v}{g} \geq  3.1~10^{-9},
    \label{noE}
\end{equation}
being independent of the atomic number density of the matter.  Note that the migration rate (and the dilution) considered here is of larger
than that observed at the Oklo site for heavy elements \cite{oklo}.

The upper limits on the anomalous helium abundance are very stringent
\cite{he} $r_{oE} \leq 10^{-19}$, and our rough estimate is ten orders of magnitude too large. Together with other problems of OBe Universe 
stipulated above, this rules out the OBe scenario. 

\section{Conclusion}\label{conclusion}
The advantages of the OHe composite-dark-matter scenario is that it is minimally related
to the parameters of new physics and is dominantly based on the effects of known atomic
and nuclear physics.
However, the full quantum treatment of this problem turns out to be rather
complicated and remains an open.

We have considered here the scenario in which such a barrier does not appear.
This leads to a significant role of inelastic reaction of OHe, and strongly modifies the main features of the OHe scenario. 
In the period of Big Bang Nucleosynthesis, when OHe is formed, it captures an additional He nucleus, so that the dominant 
form of dark matter becomes charged, recombining with electrons in anomalous isotopes of helium and heavier elements. 
Over-abundance of anomalous isotopes in terrestrial matter seems to be unavoidable in this case.

This makes the full solution of OHe nuclear physics, started in \cite{CKW2}, vital. 
The answer to the possibility of the creation of a dipole Coulomb barrier in OHe interaction with nuclei is crucial.
Without that barrier  one gets
 no suppression of inelastic reactions, in which O$^{--}$ binds with nuclei. These charged species form atoms (or ions) 
 with atomic cross sections, and that strongly suppresses their mobility in terrestrial matter, leading to their storage 
 and over-abundance near the 
 Earth's surface and oceans.

Hence, the model cannot work if no repulsive interaction appears at some distance between
OHe and the nucleus, and the solution to this open question of OHe nuclear physics is
vital for the composite-dark-matter OHe scenario.

\section*{Acknowledgements}
The research of J.R.C. and Q.W.
was supported by the Fonds de la Recherche Scientique - FNRS under grant 4.4501.05. Q.W. is also supported by the Fonds de la Recherche
Scientifique - FNRS as a research Fellow. The work by M.Kh. on initial cosmological conditions was supported by the Ministry of Education and Science of Russian Federation, project 3.472.2014/K  and his work on the forms of dark matter was supported by grant RFBR 14-22-03048.

\end{document}